\documentclass[prl,reprint,superscriptaddress,floatfix,longbibliography]{revtex4-1}

\usepackage[colorlinks=true,citecolor=blue]{hyperref}
\usepackage{amsmath}
\usepackage{amssymb}
\usepackage{amsfonts}
\usepackage{dsfont}
\usepackage{color}
\usepackage{soul}
\usepackage{nicefrac}
\usepackage[autostyle]{csquotes}
\usepackage[colorlinks=true,citecolor=blue]{hyperref}
\renewcommand{\H} {\mathcal{H}}
\usepackage{graphicx,graphics}

\begin{document} 

\title{Visualization of moir\'e magnons in monolayer ferromagnet}

\author{Somesh Chandra Ganguli}
\email{Email: somesh.ganguli@aalto.fi, jose.lado@aalto.fi, peter.liljeroth@aalto.fi}
\affiliation{Department of Applied Physics, Aalto University, FI-00076 Aalto, Finland}

\author{Markus Aapro}
\affiliation{Department of Applied Physics, Aalto University, FI-00076 Aalto, Finland}

\author{Shawulienu Kezilebieke}
\affiliation{Department of Physics, Department of Chemistry and Nanoscience Center, 
University of Jyväskyl\"a, FI-40014 University of Jyväskyl\"a, Finland}

\author{Mohammad Amini}
\affiliation{Department of Applied Physics, Aalto University, FI-00076 Aalto, Finland}

\author{Jose L. Lado}
\email{Email: somesh.ganguli@aalto.fi, jose.lado@aalto.fi, peter.liljeroth@aalto.fi}
\affiliation{Department of Applied Physics, Aalto University, FI-00076 Aalto, Finland}

\author{Peter Liljeroth}
\email{Email: somesh.ganguli@aalto.fi, jose.lado@aalto.fi, peter.liljeroth@aalto.fi}
\affiliation{Department of Applied Physics, Aalto University, FI-00076 Aalto, Finland}

\date{\today}

\begin{abstract}
%{\bf
Two-dimensional magnetic materials provide an ideal platform to explore collective many-body excitations associated with spin fluctuations. In particular, it should be feasible to explore, manipulate and ultimately design magnonic excitations in two-dimensional van der Waals magnets in a controllable way. Here we demonstrate the emergence of moir\'e magnon excitations, stemming from the interplay of spin-excitations in monolayer CrBr$_3$ and the moir\'e pattern arising from the lattice mismatch with the underlying substrate. The existence of moir\'e magnons is further confirmed via inelastic quasiparticle interference, showing the appearance of a dispersion pattern correlated with the moir\'e length scale. Our results provide a direct visualization in real-space of the
dispersion of moir\'e magnons, demonstrating the versatility of moir\'e patterns in creating emerging many-body excitations.
%} 

%We observe that the Van Hove singularities associated with the Magnonic density of states at energies 3.05, 5.75 and 9.78 meV energies disperse with out-of-plane magnetic field with slopes between $(15.35-22.6)\mu_B$. This larger than usual dispersion with magnetic field is shown to be possibly attributed to the subtrate-mediated exchange. We also demonstrate the dispersion of the magnon band and associated spin-wave excitation gap by employing quasi-particle interference technique. All these results will pave a new way for spintronic applications of the two-dimensional van der Waals magnets.  

\end{abstract}

\maketitle

The recent discovery of two dimensional van der Waals (vdW) monolayer magnetic materials has opened new avenues for scalable, defect-free samples for spintronic applications and artificial designer materials \cite{gong2017discovery,huang2017layer,burch2018magnetism,chen2018topological,chen2021magnetic,ghazaryan2018magnon,chen2019direct,kim2019evolution,reviewmag2021,mitra2021magnon}. It provides an exciting opportunity to control and manipulate magnetism in two-dimensions \cite{doping2018,electrical2018,control2018,pressure2019}, and create new emergent states in vdW heterostructures \cite{engineering2017,optical2018,TSC2020,Kezilebieke_2022_moire,HF2021}. A common feature of two-dimensional materials is the appearance of moir\'e patterns due to the lattice mismatch or twist between the monolayer and the substrate. Using the twist degree of freedom has emerged as a powerful strategy to design new quantum states \cite{PhysRevB.82.121407,PhysRevLett.99.256802,moire2013,andrei2009}. Paradigmatic examples are the emergent correlated and topological states in graphene moir\'e multilayers \cite{supertbg2018,qahtbg2020},
ferroelectricity in hexagonal boron nitride moir\'e bilayers \cite{ferroelectricity2021},
moir\'e excitons in twisted MoSe$_2$/WSe$_2$ \cite{exciton2019}, and moir\'e magnetism in CrI$_3$ moir\'e bilayers \cite{Song2021}.
The emergence of moir\'e phenomena in magnetic van der Waals materials is a newly explored field, and in particular the possibility of creating magnon moir\'e excitations remains an open problem in twistronics.

\begin{figure*}[t!]
    \centering
    \includegraphics[width=.9\textwidth]{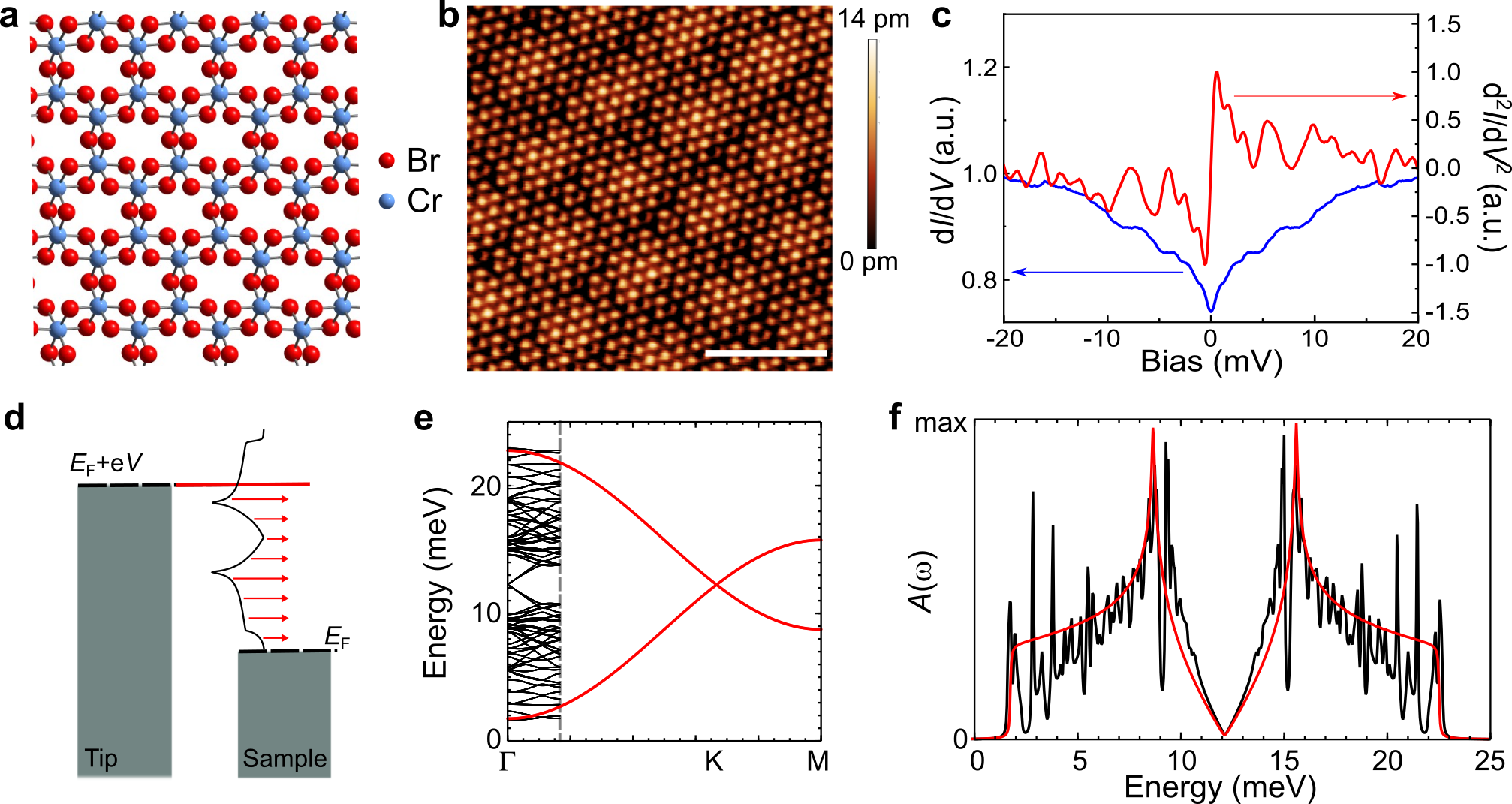}
	\caption{\textbf{Probing moir\'e magnons in CrBr$_3$ with IETS.} \textbf{a}, Schematic of the CrBr$_3$ structure. \textbf{b}, Atomically resolved image of CrBr$_3$ on HOPG. Image was taken at sample bias 1.5 V. Scale bar 5 nm. \textbf{c}, Averaged and symmetrized d$I/$d$V$ (blue) and numerically differentiated d$^{2}I/$d$V^{2}$ (red) obtained in monolayer CrBr$_3$ on HOPG. \textbf{d}, Schematic of inelastic tunneling spectroscopy of magnons. \textbf{e}, Unfolded magnon bands and moir\'e-folded magnon mini-bands. \textbf{f}, Unfolded (red) and moir\'e-folded (black) magnonic spectral functions. }
	    \label{fig:fig1}
    \label{fig:sketch}
\end{figure*}

%Chromium trihalides (CrX$_3$, X= Cl, Br and I, Fig.~\ref{fig:fig1}a) have been established as a prominent family of 2D magnetic materials \cite{halides2017} with all three showing ferromagnetic order, where the easy axis is out-of-plane for CrBr$_3$ \cite{ghazaryan2018magnon,chen2019direct} and CrI$_3$ \cite{huang2017layer}, and in-plane for CrCl$_3$ \cite{CrCl32021}. Here, we demonstrate the emergence of moir\'e magnons in CrBr$_3$, arising from a moir\'e pattern with underlying substrate. We show that the interplay between magnetic quantum fluctuations of the ferromagnet and the underlying substrate results in a reconstruction of the magnon dispersion, leading to new van Hove singularities in the magnon spectral function. We probe the moir\'e magnon spectra by means of inelastic scanning tunneling spectroscopy on monolayer CrBr$_3$ under out-of-plane external magnetic field, and demonstrate a one-to-one correlation between the moir\'e length scale and the magnon van Hove singularities. Furthermore, by exploiting quasiparticle interference with inelastic spectroscopy, we directly probe the magnon dispersion in reciprocal space, allowing us to map the moir\'e magnon spectra. Our results demonstrate the emergence of moir\'e magnons and the impact of moir\'e patterns on the magnetic excitations of 2D materials.
%The system we present consists of a CrBr$_3$ monolayer on top of HOPG (sample preparation described in the Methods section). 

Chromium trihalides (CrX$_3$, X= Cl, Br and I, Fig.~\ref{fig:fig1}a) have been established as a prominent family of 2D magnetic materials \cite{halides2017} with all three showing ferromagnetic order, where the easy axis is out-of-plane for CrBr$_3$ \cite{ghazaryan2018magnon,chen2019direct} and CrI$_3$ \cite{huang2017layer}, and in-plane for CrCl$_3$ \cite{CrCl32021}. We have carried out low-temperature scanning tunneling microscopy (STM) and spectroscopy (STS) to probe the magnon excitations in monolayer CrBr$_3$. We show that the results can be understood in terms of moir\'e magnons arising from a reconstruction of the magnon dispersion by the moir\'e pattern formed by the lattice mismatch between CrBr$_3$ and the substrate. This leads to new van Hove singularities in the magnon spectral function that are correlated with the moir\'e length scale. Furthermore, by exploiting quasiparticle interference with inelastic spectroscopy, we directly probe the magnon dispersion in reciprocal space, allowing us to map the moir\'e magnon spectra. Our results demonstrate the emergence of moir\'e magnons and the impact of moir\'e patterns on the magnetic excitations of 2D materials.

We have carried out experiments on CrBr$_3$ monolayers on a highly-oriented pyrolytic graphite (HOPG) substrate at a $T=350$ mK (see Methods section for more experimental details). Typical STM topography image (Fig.~\ref{fig:fig1}b) shows both bright triangular protrusions arising from the bromine atoms in the CrBr$_3$ layer as well as a longer length-scale variation corresponding to the moir\'e pattern, which arises from the lattice mismatch between the CrBr$_3$ monolayer and the HOPG substrate. Magnetic excitations can be probed via inelastic tunneling spectroscopy (IETS) %as a function of magnetic field  
and they should result in bias-symmetric steps in the d$I$/d$V$ signal \cite{spinelli2014imaging,Ternes_2015,tunneling2018,ghazaryan2018magnon}. We observe clear inelastic excitations experimentally as demonstrated in Fig.~\ref{fig:fig1}c that shows both the measured d$I$/d$V$ (symmetrized) and numerically differentiated and smoothened d$^2I$/d$V^2$ signals (see Supplementary Information (SI) for details). As schematically illustrated in Fig.~\ref{fig:fig1}d, 
%\SK{consider to replace this with field dependence IETS data in SI} \JL{The LL of the magnetic field dependence makes the spectra a bit hard to interpret, on my side perhaps it fits better in the SI}
for a ferromagnetic system, we would expect the d$I$/d$V$ to correspond to the integrated magnon DOS while the d$^2I/$d$V^2$ signal directly corresponds to the local magnon spectral function. It is immediately obvious that our experimental d$^2I/$d$V^2$ contains many more peaks than expected for a typical magnon spectrum. We explain this discrepancy below as arising from the moir\'e-induced modification of the magnon spectrum.

 The physics behind the moir\'e magnons can be understood starting from the anistropic Heisenberg Hamiltonian \cite{anisotropy2017,morpurgo2019} describing the spin excitations in a magnetic two-dimensional system (see SI for details)

\begin{equation}
    \H = -\sum_{ij} J_{ij} \vec S_i \cdot \vec S_j 
    -\sum_{ij} K_{ij} S^z_i S^z_j 
    +\mathcal{H}_V
\end{equation}

with $J_{ij}$ the spatially modulated isotropic exchange coupling, $K_{ij}$ the anisotropic exchange and $S^\alpha_n$ the local $S=3/2$ operators in the Cr atoms, \footnote{We note that additional terms can be included in the Hamiltonian, leading to analogous results \cite{anisotropy2017,chen2018topological,Fazel2020,morpurgo2019}.} forming a honeycomb lattice (Figs.~\ref{fig:fig1}a). The term $\mathcal{H}_V$ contains other potential terms in the Hamiltonian including Dzyaloshinskii-Moriya interaction, biquadratic exchange, single-ion anisotropy and Kitaev interaction, which for the sake of simplicity are not included in the next discussion as their role is not important for the emergence of moir\'e magnons. The local moments at the Cr-sites have a ferromagnetic coupling via superexchange through Br atom, parametrized by $J_{ij}$. The existence of the substrate leads to an additional exchange interaction mediated by the RKKY interaction. This substrate-mediated RKKY interaction depends on the local stacking between CrBr$_3$ and HOPG, which in turn is controlled by the moir\'e between HOPG and CrBr$_3$. This modulation in real space leads to the change of the exchange constants $J_{ij}$ \cite{chen2019direct,stacking2018,Song2021,2021arXiv210309850X} and, in turn, the spin stiffness through the moir\'e unit cell \cite{PhysRevB.102.094404,PhysRevLett.125.247201,2022arXiv220605264K,moiremagnon2022}. Moreover, potential small structural distortions lead to a modulation of the superexchange interaction, both of which follow the same periodicity as the moir\'e pattern. Holstein-Primakoff mapping \cite{PhysRev.58.1098} allows the magnonic Hamiltonian to be written in terms of the bosonic magnon operators
\begin{equation}
    \H = 
    -\sum_{ij}\gamma_{ij} a^\dagger_i a_j 
    + \sum_n \Delta_n a^\dagger_n a_n+ \text{h.c.}
\end{equation}
with $\gamma_{ij} \sim J_{ij}$ controlling the spin stiffness and $\langle \Delta_n \rangle $ determines the magnon gap, and $a^\dagger_n, a_n$ are the creation and annhilation magnon operators. For CrBr$_3$, first principles calculations \cite{Zhang2015} predict a bandwidth of the magnon spectra of $\sim 30$ meV in the absence of a moir\'e pattern, and in the following we take that the moir\'e modulations changes the local exchange while keeping the global bandwidth approximately equal to the uniform case.

In the absence of the moir\'e pattern, the magnon dispersion features two magnon bands stemming from the two Cr atoms in the unit cell. The magnon dispersion shows Dirac points when neglecting small contributions coming from $\mathcal{H}_V$, and a low energy quadratic dispersion with a gap controlled by $K$. In the presence of the moir\'e pattern, the real-space modulation of $\gamma_{ij}$ leads to the appearance of magnon mini-bands in the moir\'e supercell, as shown in Fig.~\ref{fig:fig1}e. The multiple folding of the original moir\'e bands and induced anticrossings driven by the moir\'e exchange modulation gives rise to a whole new set of moir\'e singularities, as shown in Fig.~\ref{fig:fig1}f.

\begin{figure*}[t!]
    \centering
    \includegraphics[width=.9\textwidth]{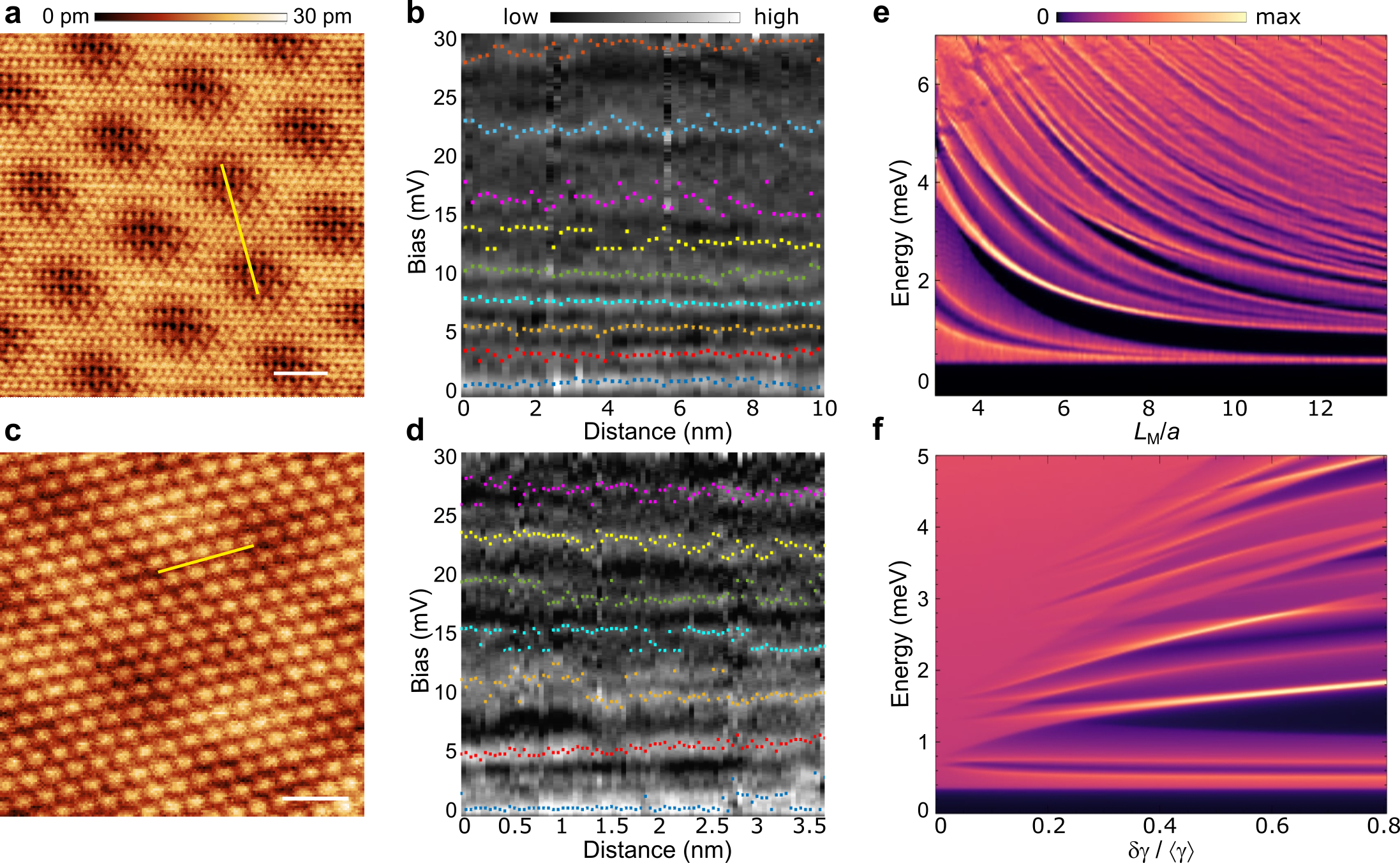}
	\caption{\textbf{Moir\'e magnons in CrBr$_3$.} \textbf{a,b}, Area with moir\'e wavelength 7 nm (Scale bar 4 nm. Image bias 1 V), spatial dependence of antisymmetrised $d^{2}I/dV^{2}$ (b) with spectra taken along yellow line in panel (a). \textbf{c,d} Area with moir\'e wavelength 3.6 nm (Scale bar 2 nm. Image bias 2 V), spatial dependence of antisymmetrised $d^{2}I/dV^{2}$ (d) with spectra taken along yellow line in panel (c). In (b),(d), colored points indicate the locations of maxima in the $d^{2}I/dV^{2}$ corresponding to the van Hove singularities in the magnon spectral function (see SI). \textbf{e,f} Theoretical dependence of the magnon spectral function with the moir\'e length (e) and with the strength of the exchange modulation (f).}
    \label{fig:moire}
\end{figure*}

We have carried out inelastic tunneling spectroscopy experiments (parameters mentioned in the Methods section) over a range of CrBr$_3$ islands with different moir\'e periodicities allowing us to address the impact of the underlying moir\'e pattern on the magnetic excitations. Figs.~\ref{fig:moire}a,c show the effect of the  different moir\'e length scales (moir\'e wavelengths of 7 and 3.6 nm) on the inelastic excitations. The anti-symmetrised d$^{2}I/$d$V^{2}$ (details in the SI) shows strong peaks that are spatially quite uniform as shown in Fig.~\ref{fig:moire}b,d. %\PL{How to write the next sentence in a reasonable way? Maybe compare the peaks like: appear at energies of 1.08 (0.6) meV, 3.6 (1.68) meV, ..., respectively.} The inelastic tunneling peaks appear at average energies 1.08 meV, 3.6 meV, 5.88 meV, 7.92 meV, 16.68 meV, 19.2 meV, 22.08 meV, 23.04 meV, 25.08 meV and and at average energies 0.6 meV, 1.68 meV , 3.12 meV, 5.4 meV, 9.84 meV, 11.64 meV, 13.56 meV, 15.48 meV, 17.88 meV, 19.44 meV, 22.8 meV, 25.56 meV, 26.88 meV for the areas with moir\'e wavelengths 7 nm and 3.6 nm respectively (see supplementary information). 
The moir\'e magnon features are expected to be the most visible at the bottom of the magnon band due to the folding into the moir\'e Brillouin zone (see Fig.~\ref{fig:fig1}f). In addition, the higher energy inelastic excitations are less intense in the experimental spectra, which can be understood through magnon-magnon interation effects \cite{PhysRevX.8.011010}. Therefore, we focus on the lower energy features and comparing Figs.~\ref{fig:moire}c,d, it is clear that there are more inelastic features with a smaller energy spacing in the experiments on the larger length-scale moir\'e pattern (see SI for the statistics of inelastic peak energies). 

The dependence of low energy inelastic magnon peaks with the moir\'e wavelengths can be rationalized from the reconstruction of the magnon bands triggered by the moir\'e pattern. The momentum folding of the magnon structure depends on the length of the moir\'e pattern, leading to magnon van Hove singularities whose energy location depends on the specific moir\'e. In particular, longer moir\'e lengths give rise to magnon van Hove singularities with a smaller energy spacing, as observed experimentally. This phenomenology is captured with the moir\'e Heisenberg model, as shown in Fig.~\ref{fig:moire}e. %Moir\'e reconstruction leads to a low energy shifting of the magnon van Hove singularities, as observed experimentally.\PL{previous sentence is not English. Shorther wavelength moir\'e reconstruction leads to a shift of the magnon van Hove singularities to lower energies, as observed experimentally?}
The relative intensity of the moir\'e van Hove singularities
is controlled by the strength of the moir\'e modulation
as shown in Fig.~\ref{fig:moire}f, highlighting that the observation of moir\' e magnons can allow inferring the value of the real space modulation of the exchange constants.

\begin{figure*}[t!]
    \centering
    \includegraphics[width=.9\textwidth]{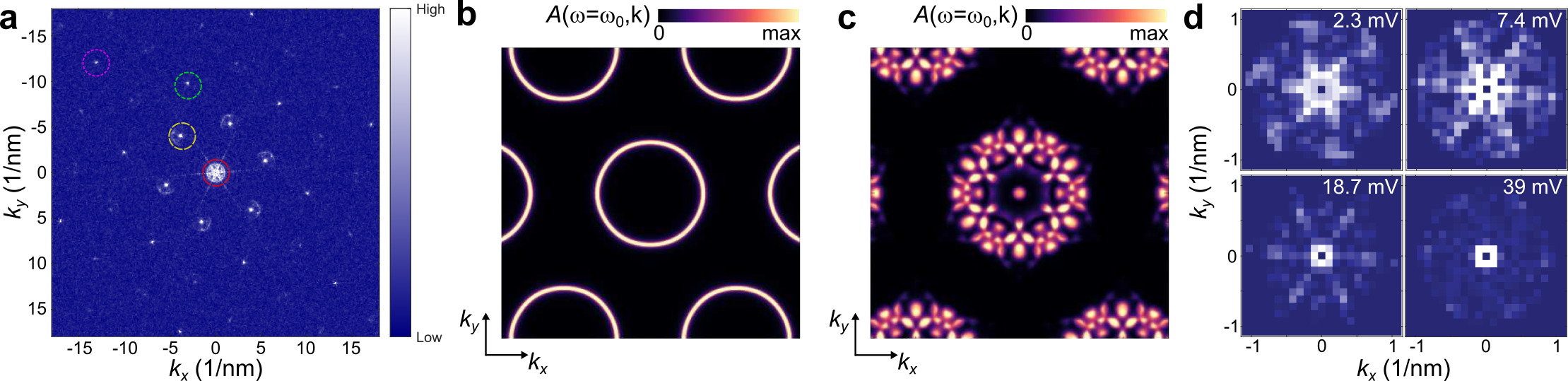}
	\caption{\textbf{Quasiparticle interference of magnons.} \textbf{a}, FFT of a constant-current d$I/$d$V$ map at a bias voltage of 7.4 mV. Red, yellow, green and magenta dotted circles indicate real space lengthscales of 7 nm, 1.25 nm, 6 {\AA}, 4  {\AA},   respectively. 
	\textbf{b,c}, Calculated momentum-resolved spectral function of magnon $A(\omega,\mathbf k)$ at constant energy of 5 meV in the absence (b) and presence of the moir\'e pattern (c). \textbf{d}, Zoomed-in FFTs of the experimental d$I$/d$V$ signal around the $\Gamma$-point at the bias voltages indicated in the panels.}
    \label{fig:qpi}
\end{figure*}

While our experiments are consistent with the expectation that the observed inelastic features correspond to magnetic excitations, they could also correspond to inelastic excitations of phonons. However, earlier experiments on tunneling devices have shown that the modes with sufficient electron-phonon coupling are at higher energies (above 25 meV) \cite{Ghazaryan2018} than the features we observe in our experiments. Additionally, the magnetic origin of the excitations is usually probed by carrying out experiments under an external magnetic field. We have done these experiments (see SI for the results); however, the HOPG substrate shows very clear and strong signatures of Landau levels at high magnetic fields that completely overwhelm the signal from the magnetic excitations in the CrBr$_3$ layer. It is also not possible to subtract the signal from the Landau levels as the Landau level spectra of bare HOPG and HOPG covered by CrBr$_3$ are different arising from the sensitivity of the Landau levels to the local potential \cite{Morgenstern2001Landau}. At low-magnetic fields ($<0.5$ T), the signal due to the magnetic excitations is clearly visible, but the shifts due to the Zeeman energy are too small to be reliably detected.

The presence of moir\'e magnons can be demonstrated even more convincingly through inelastic quasiparticle interference spectroscopy that allows a direct visualization of the length scale of the moir\'e magnons. We note that this technique has been used to demonstrate the emergence of quantum spin liquid signatures in monolayer 1T-TaSe$_2$ \cite{spinon2021}. The differential conductance d$I/$d$V$ is proportional to the total number of magnons that can be excited with that energy. In the presence of weak scattering, the total number of magnons will be spatially modulated. The dispersion of the moir\'e magnons can be directly probed by visualizing the Fourier transform of the spatially resolved d$I/$d$V$, shown in Fig.~\ref{fig:qpi}a, known as quasiparticle interference (QPI). The signature of moir\'e magnons is directly visible in the QPI due to the reconstruction of the magnon spectra. Specifically, the Fourier transform of the d$I/$d$V$, in the following denoted as $\Xi (\omega,\mathbf q)$ stems from inelastic magnon tunneling processes as $\Xi (\omega,\mathbf q)\sim 
\int A(\omega,\mathbf k) A(\omega,\mathbf k + \mathbf q) d^2\mathbf{k} $, where $A(\omega,\mathbf k)$ is the magnon spectral function.
As a result, the magnon QPI reflects a self-convolution of the
magnon dispersion, directly reflecting magnon reconstructions in
reciprocal space. 

While this kind of QPI features could also arise from elastic scattering between electronic states, it is very unlikely in the present case. First of all, CrBr$_3$ is an insulator and has no electronic states close to the Fermi level. We could of course still in principle observe QPI from the electronic states of the HOPG substrate; however, in that case one would expect QPI signal over a large bias range since HOPG has states at all energies. This is in contrast to our experimental results and hence, the QPI features most likely correspond to the magnon excitations.

To explore the dispersion of the magnonic bands, we performed constant current d$I/$d$V$ maps at various energies (parameters mentioned in the Methods section). The typical FFT of the d$I/$d$V$ maps has strong peaks at characteristic reciprocal space points, indicating different topographic periodicities present. The green, magenta, red and yellow dotted circles in Fig.~\ref{fig:qpi}a represents Cr-Cr (6  {\AA}), Br-Br (4  {\AA}), moir\'e (7 nm) and possible $\sqrt{3} \times \sqrt{3}$ Kekul\'e distortion (1.25 nm) length scales. The high-symmetry points, especially $\Gamma$ and $K$ points have features around them.

Theoretically, in the absence in the moir\'e pattern, 
the magnon spectral function at energies below 8 meV
should feature a simple circular shape coming from the magnon dispersion $\epsilon(\mathbf k) \sim |\mathbf{k}|^2$
as shown in Fig.~\ref{fig:qpi}b. This featureless circular shape leads to the well-known disc-like QPI, that does not show a complex angular structure. In stark contrast, in the presence of the moir\'e,
the moir\'e modulation leads to a full new set
features in the magnon dispersion as shown in Fig.~\ref{fig:qpi}c, 
as a direct consequence of the magnon moir\'e mini-bands. 
The inelastic contribution to the QPI gives rise to the different scattering events associated with the states in Fig.~\ref{fig:qpi}c, directly reflecting the emergent dispersion of the moir\'e bands. In particular, the moir\'e magnon generating QPI will give rise to very short wavelength features appearing around $\Gamma$ point in the QPI.

These theoretical moir\'e QPI predictions can be directly compared with our experimental data (Fig.~\ref{fig:moire}c).
In order to factor out the impact of the topographic moir\'e modulation in the QPI, we
first remove the peaks associated with the moir\'e length, whose origin is purely structural.
Around the $\Gamma$ point, after removing the intensity due to the moir\'e, we see an internal
interference pattern strongly dependent on the energy and ultimately vanishing above $\sim25$ mV (Fig.~\ref{fig:qpi}d). 
It must be noted that, in the absence of a moir\'e pattern, no strong energy dependence of the QPI is expected around the $\Gamma$ point. In stark contrast, the presence of moir\'e magnons leads
to an energy-dependent interference pattern around the $\Gamma$ point in the full
energy window due to the non-trivial interplay between the different magnon moir\'e bands.
The previous phenomenology directly demonstrates the emergence of quasiparticle interference
associated with magnons, featuring fluctuations in the moir\'e length scale and spanning over the whole
energy window in which magnons fluctuations appear in CrBr$_3$.
%Around K-point, we see a feature persisting upto ~20 mV. We also observe thermal smearing of the magnon band structure (see supplementary material).

To summarize, we have demonstrated the emergence of moir\'e magnon excitations in 2D monolayer ferromagnet. By using inelastic spectroscopy, we showed that the existence of moir\'e patterns with different moir\'e lengths leads to different reconstructions of the moir\'e spectra. The existence of moir\'e magnons is further confirmed via inelastic quasiparticle interference, showing the appearance of a dispersion pattern correlated with the moir\'e length scale. Our results provide a direct visualization in real space of the dispersion of moir\'e magnons, demonstrating the versatility of moir\'e patterns in creating emergent many-body excitations.

%The presence of moir\'e magnon and strong Landau level signal coming from HOPG renders the estimation of magnon dispersion with external magnetic field difficult. In future, we propose to create a barrier between HOPG substrate and CrBr$_3$ to reduce these effects.

\section{Methods}

\textbf{Sample growth:} The CrBr$_3$ thin film was grown on freshly cleaved HOPG substrates by compound source molecular beam epitaxy. The anhydrous CrBr$_3$ flakes of 99 \% purity was evaporated by Knudsen cell. Before growth, the cells were degassed up to the growth temperature $350^\circ$C until the vacuum was better than $1\times10^{-8}$ mbar. The growth rate was determined by checking the coverage of the as-grown samples by STM.

\textbf{STM measurements:} Subsequent to the growth, the sample was transferred to a low-temperature STM (Unisoku USM-1300) housed in the same UHV system. STM imaging and STS experiments were performed at $T=350$ mK. STM imaging was performed in constant current mode. Differential conductance (d$I$/d$V$) spectra were measured using standard lock-in techniques sweeping the sample bias in an open feedback loop with a.c bias modulation at a frequency of 873.7 Hz. For the d$I$/d$V$ maps in Fig.~2b,d, the amplitude of bias modulation was 500 $\mu$V and the current set point was 500 pA. For constant current d$I$/d$V$ maps in Fig.~3a,d, the amplitude of bias modulation was kept to 5\% of the applied d.c. bias, and the current set point was 200 pA. The raw images were drift corrected by Lawler-Fujita algorithm \cite{lawler2010intra} and symmetrized to increase the signal to noise ratio of the QPI signal (described in SI).

%\section{Data availability}
%The datasets generated during and/or analysed during the current study are available from the corresponding authors on reasonable request.

\bibliography{biblio}

\section{Acknowledgments}
This research made use of the Aalto Nanomicroscopy Center (Aalto NMC) facilities and was supported by the European Research Council (ERC-2017-AdG no.~788185 ``Artificial Designer Materials'') and Academy of Finland (Academy professor funding nos.~318995 and 320555, Academy research fellow no.~331342 and 336243), and the Jane and Aatos Erkko Foundation. We acknowledge the computational resources provided by the Aalto Science-IT project.

%\section{Author contributions}
%SCG, JLL, PL conceptualised the problem. SCG, MA, SK did the sample growth, SCG, SK performed the STM experiments, SCG, MA, MhA did the data analysis, JLL performed the theoretical calculations, SCG, JLL, PL wrote the manuscript. All the authors commented on the manuscript.

\section{Competing interests}
The authors declare no competing interests.

\end{document}

\title{Supplementary Information: Visualization of moir\'e magnons in monolayer ferromagnet}

\author{Somesh Chandra Ganguli}
\email{Email: somesh.ganguli@aalto.fi, jose.lado@aalto.fi, peter.liljeroth@aalto.fi}
\affiliation{Department of Applied Physics, Aalto University, FI-00076 Aalto, Finland}

\author{Markus Aapro}
\affiliation{Department of Applied Physics, Aalto University, FI-00076 Aalto, Finland}

\author{Shawulienu Kezilebieke}
\affiliation{Department of Physics, Department of Chemistry and Nanoscience Center, 
University of Jyväskyl\"a, FI-40014 University of Jyväskyl\"a, Finland}

\author{Mohammad Amini}
\affiliation{Department of Applied Physics, Aalto University, FI-00076 Aalto, Finland}

\author{Jose L. Lado}
\email{Email: somesh.ganguli@aalto.fi, jose.lado@aalto.fi, peter.liljeroth@aalto.fi}
\affiliation{Department of Applied Physics, Aalto University, FI-00076 Aalto, Finland}

\author{Peter Liljeroth}
\email{Email: somesh.ganguli@aalto.fi, jose.lado@aalto.fi, peter.liljeroth@aalto.fi}
\affiliation{Department of Applied Physics, Aalto University, FI-00076 Aalto, Finland}

%\begin{abstract}
%Here we give details about the theoretical model and the magnetic field dependence of the magnon bands.
%\end{abstract}

%\date{\today}

\maketitle

\section{Determining the inelastic spectral function}

\begin{figure}[h!]
    \centering
    \includegraphics[width=.95\columnwidth]{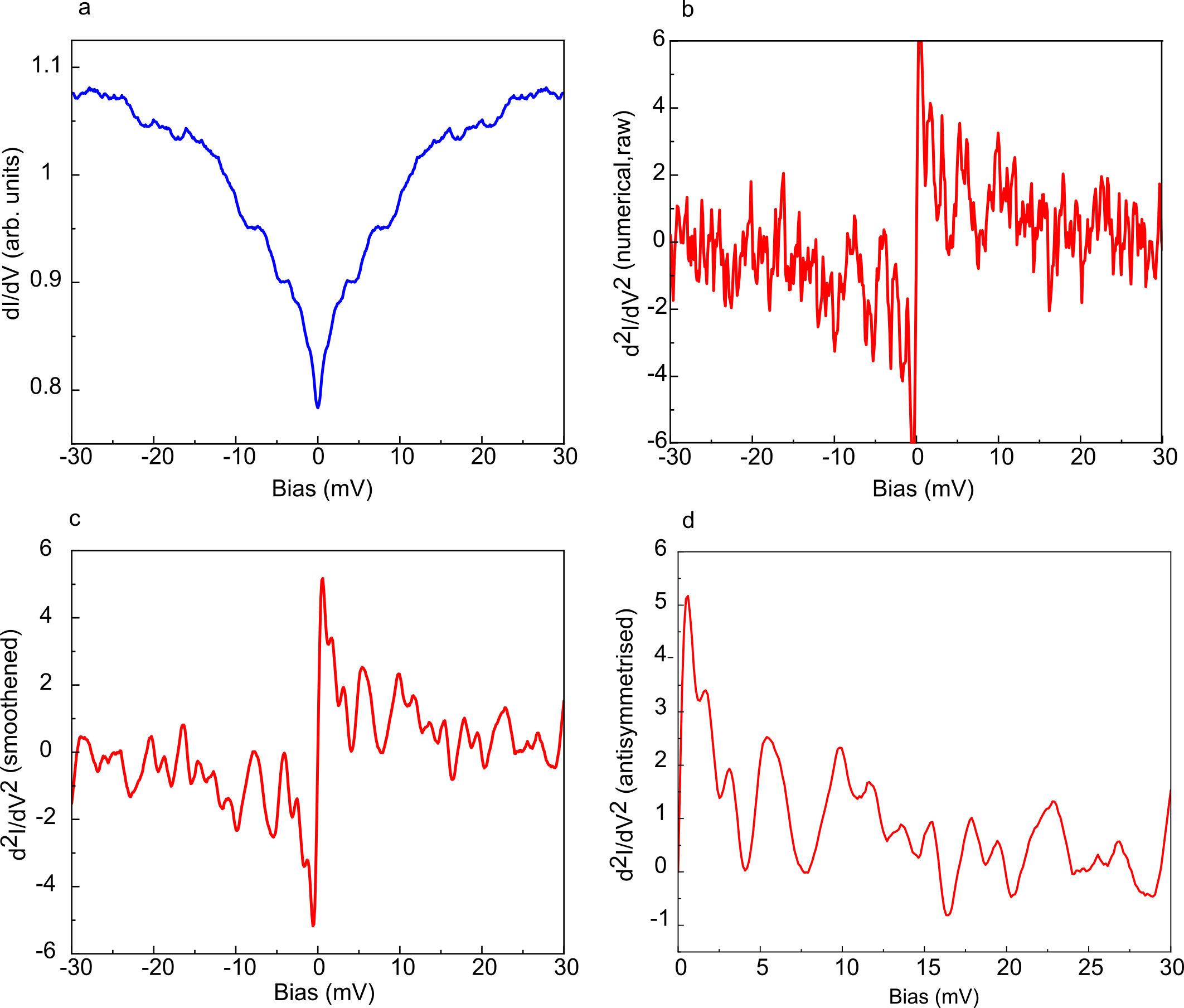}
	\caption{Numerically determined inelastic spectral function: (a) Experimental d$I$/d$V$, (b) numerically differentiated d$^{2}I/$d$V^{2}$, (c) smoothened d$^{2}I/$d$V^{2}$, and (d) antisymmetrised d$^{2}I/$d$V^{2}$.}
    \label{figSM:Inel}
\end{figure}

The experimental d$I$/d$V$ signal (Fig.~\ref{figSM:Inel}a) is numerically differentiated to obtain d$^{2}I/$d$V^{2}$ (Fig.~\ref{figSM:Inel}b). The numerical d$^{2}I/$d$V^{2}$ is smoothened by Savitzky–Golay method to obtain smoothened d$^{2}I/$d$V^{2}$ (Fig.~\ref{figSM:Inel}c) and the antisymmetrised d$^{2}I/$d$V^{2}$ is determined as (d$^{2}I/$d$V^{2}(V)$-d$^{2}I/$d$V^{2}(-V)$)/2.

\section{Histogram of spatially dependent inelastic excitations}

\begin{figure}[h!]
    \centering
    \includegraphics[width=.95\columnwidth]{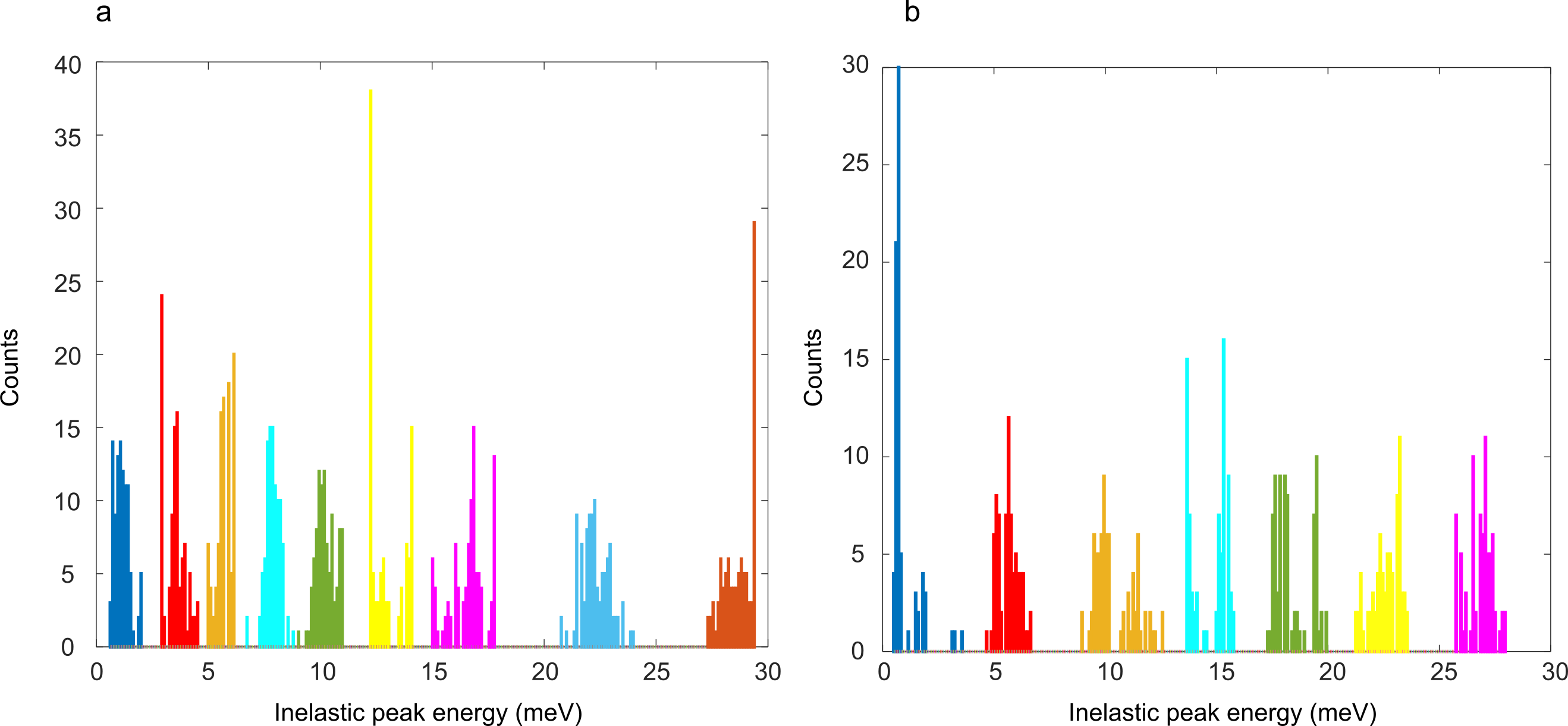}
	\caption{Histogram of inelastic excitation energies corresponding to the (a) 7 nm moir\'e area, (b) 3.6 nm moir\'e area. The colors refer to individual miniband energies for the corresponding moir\'e areas in Figs. 2b, d.}
    \label{figSM:hist}
\end{figure}

We determined the energies of the inelastic peaks corresponding to the two different moir\'e areas in Figs. 2b, d. We observe that, specific number of inelastic peaks appear throughout albeit with some spatial variations in the same moir\'e area from the histogram plot of the energy values at different spatial locations (Fig.~\ref{figSM:hist}a,b). However, the number of inelastic peaks and their energies vary in two different moir\'e areas, with the larger moir\'e area having larger number of peaks (Fig.~\ref{figSM:hist}a,b). 

%\section{Interplay between Landau levels and magnon peaks}
\section{Magnetic field dependence of the inelastic excitations}
\begin{figure}[t!]
    \centering
    \includegraphics[width=.95\columnwidth]{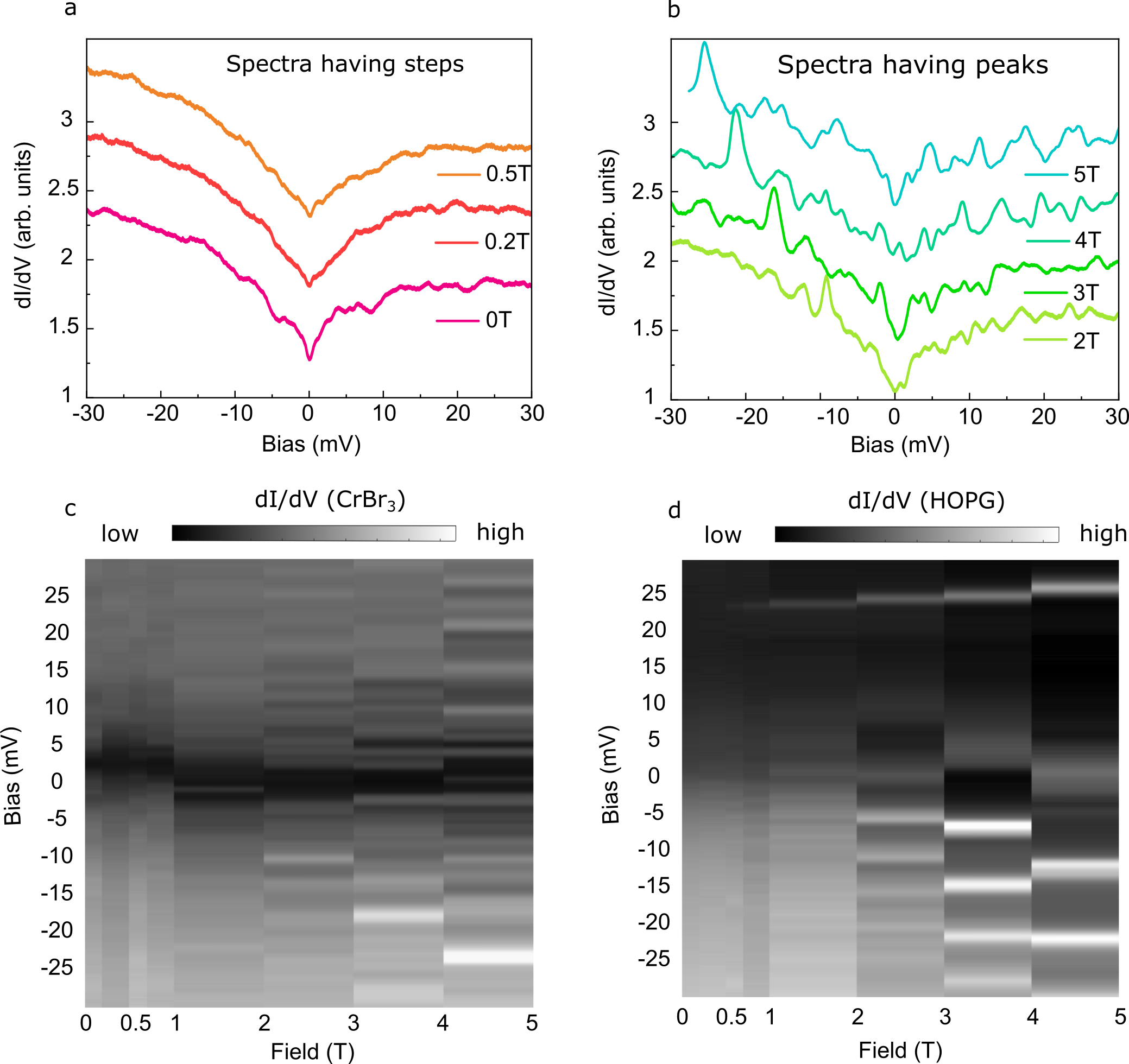}
	\caption{(a, b) Magnetic field dependence of the differential conductance (d$I$/d$V$) on CrBr$_{3}$ showing step-like features in low magnetic fields (a) and peak-like features in high magnetic fields (b). (c) Full Magnetic field dependence of (d$I$/d$V$) on CrBr$_{3}$. (d) Magnetic field dependence of the differential conductance (d$I$/d$V$) on HOPG showing the appearance of Landau levels.}
    \label{figSM:Field}
\end{figure}

When an external out-of-plane magnetic field is applied, the van Hove singularities in the magnonic band are expected to disperse linearly with the magnitude of the applied field. However, with increasing magnetic field, we observe a change of d$I$/d$V$ signal from having step features to peak-like features (Fig.~\ref{figSM:Field}a, b, c). In the presence of an external magnetic field, the substrate HOPG shows strong peak-like features in d$I$/d$V$ (Fig.~\ref{figSM:Field}d). They correspond to Landau levels (LL) due to quasi-two dimensional nature of conduction electrons within HOPG \cite{li2007observation,matsui2005sts}. These strong signals in the d$I$/d$V$ channel from the HOPG increases the number of peak like features observed in CrBr$_3$ and it makes it difficult to quantify the shifts of the magnetic excitations at high magnetic fields.

\begin{figure}[t!]
    \centering
    \includegraphics[width=.95\columnwidth]{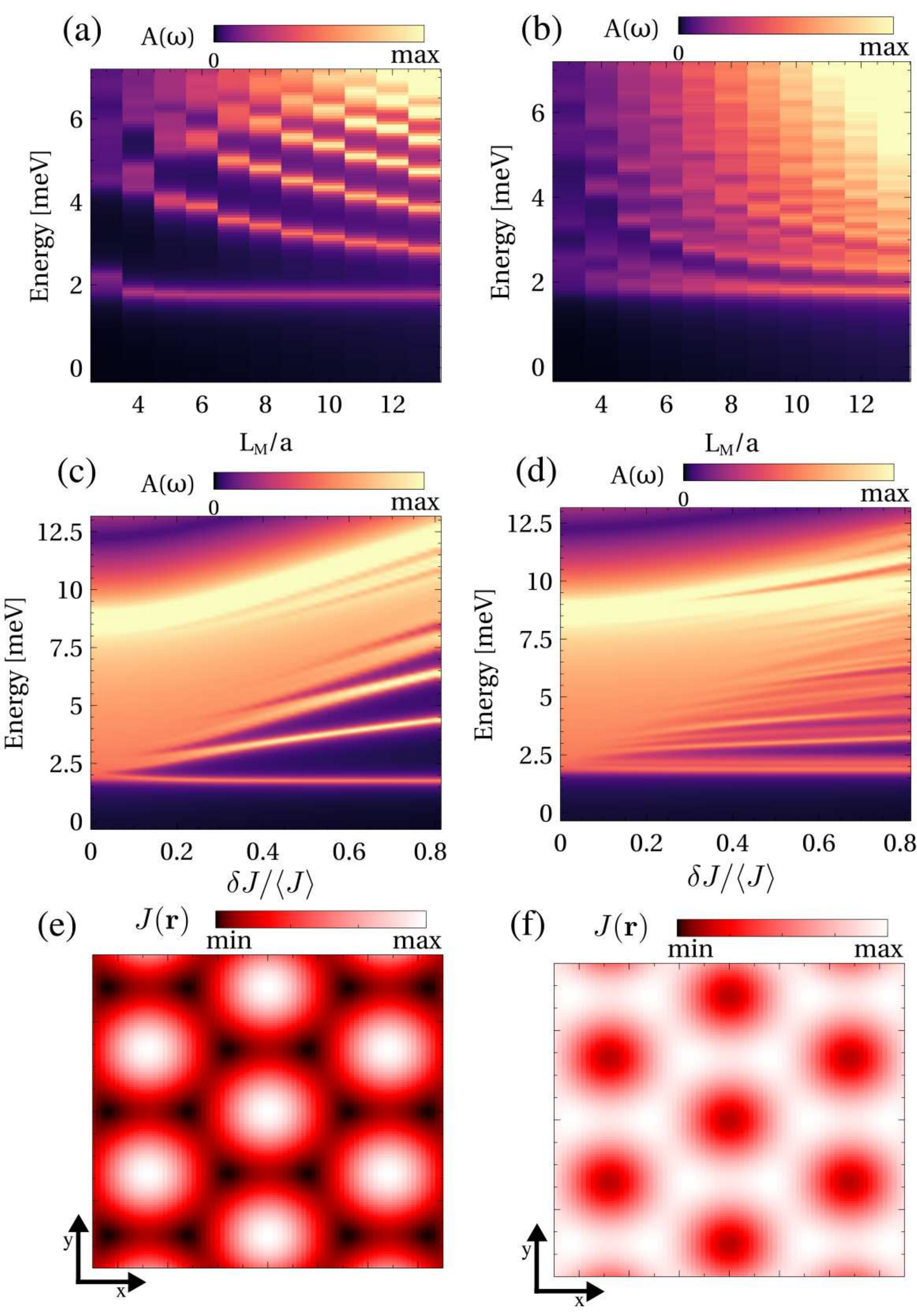}
	\caption{(a,b) Magnon spectral function
	as a function of the strength of the moir\'e potential
	and (b,c) as a function of the moir\'e length.
	Panels (a,c) and (b,d) correspond to different
	exchange modulations, as shown in panels
	(e,f), respectively.}
    \label{figSM:moire}
\end{figure}

\begin{figure}[h!]
    \centering
    \includegraphics[width=.95\columnwidth]{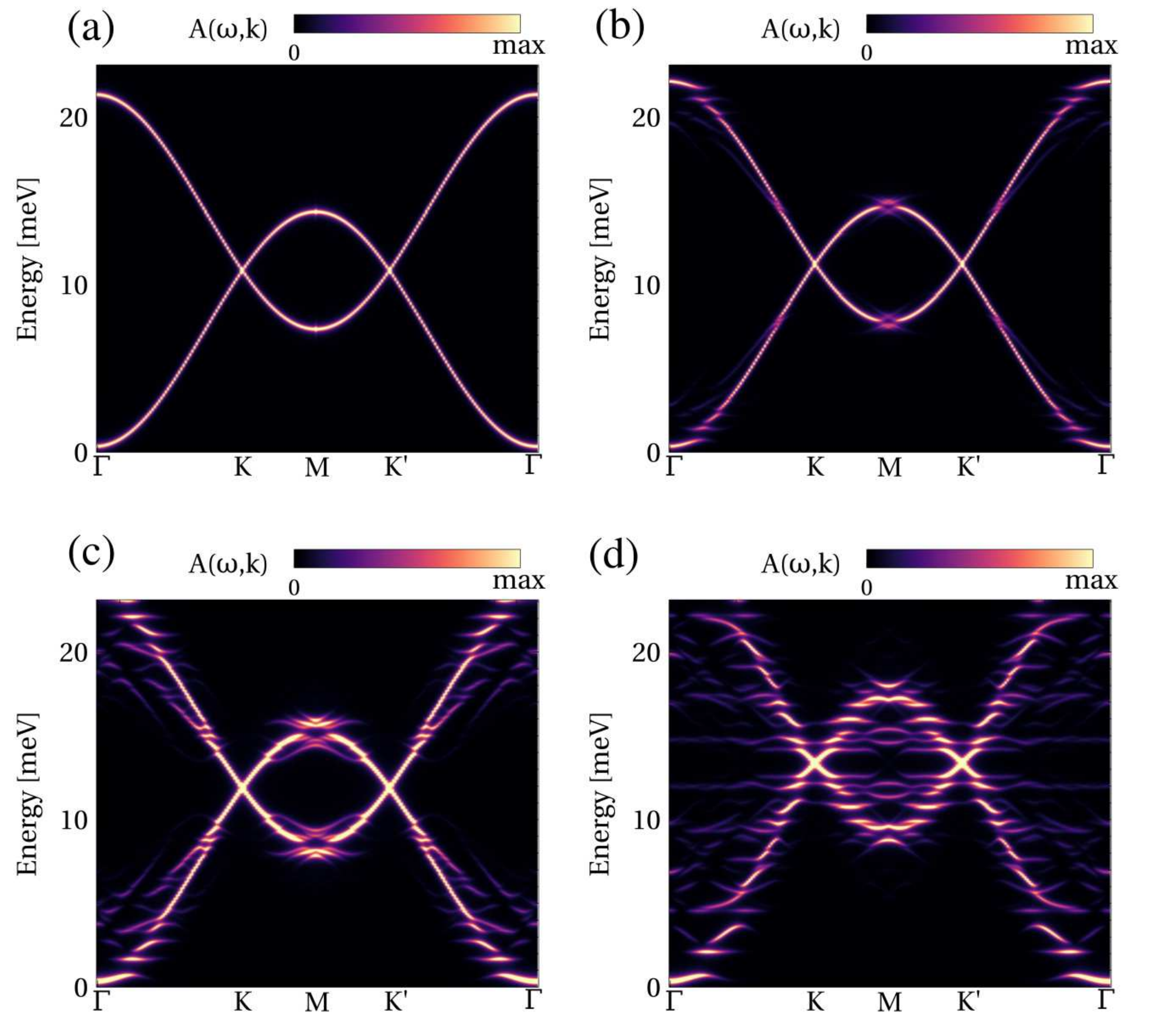}
	\caption{(a-d) Magnon spectral function unfolded
	to the minimal unit cell, for strength modulation
	$\delta J=0 $ (a)
	$\delta J=0.3 \langle J\rangle $ (b),
	$\delta J=0.6 \langle J\rangle $ (c),
	$\delta J=1.2 \langle J\rangle $ (d).
}
    \label{figSM:unfolded}
\end{figure}

\begin{figure}[b!]
\centering
%    \begin{minipage}{0.45\textwidth}
    \includegraphics[width=0.91\columnwidth]{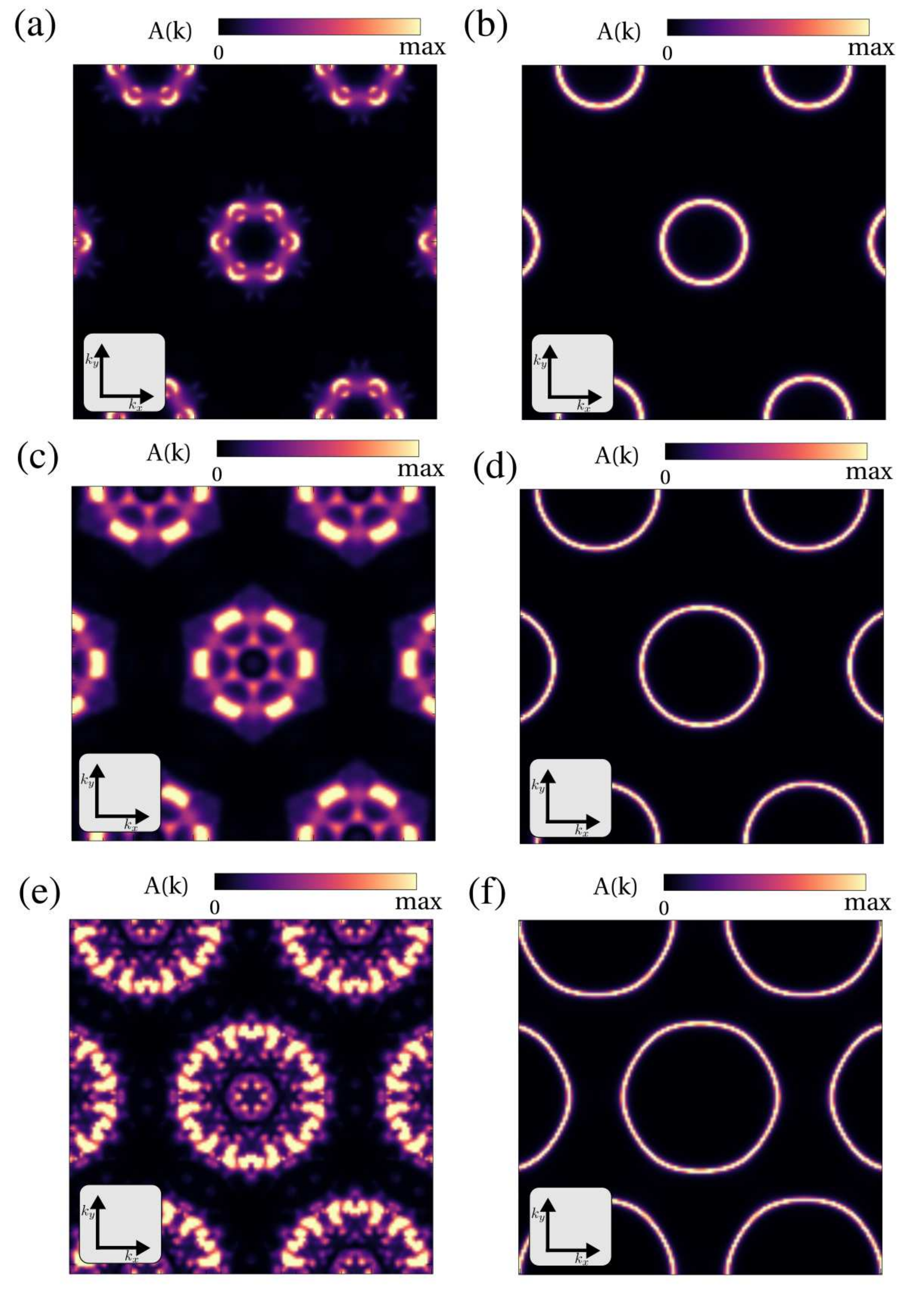}
	\caption{Unfolded magnon spectral function at a fixed energy 
	with (a,c,e) and without (b,d,f) the moir\'e exchange modulation.
	In the presence of the moire, the emergence of features close to the $\Gamma$
	point is observed in the whole energy range.} \label{figSM:FS}
\end{figure}

\begin{figure}[t!]
\centering
    \includegraphics[width=0.95\columnwidth]{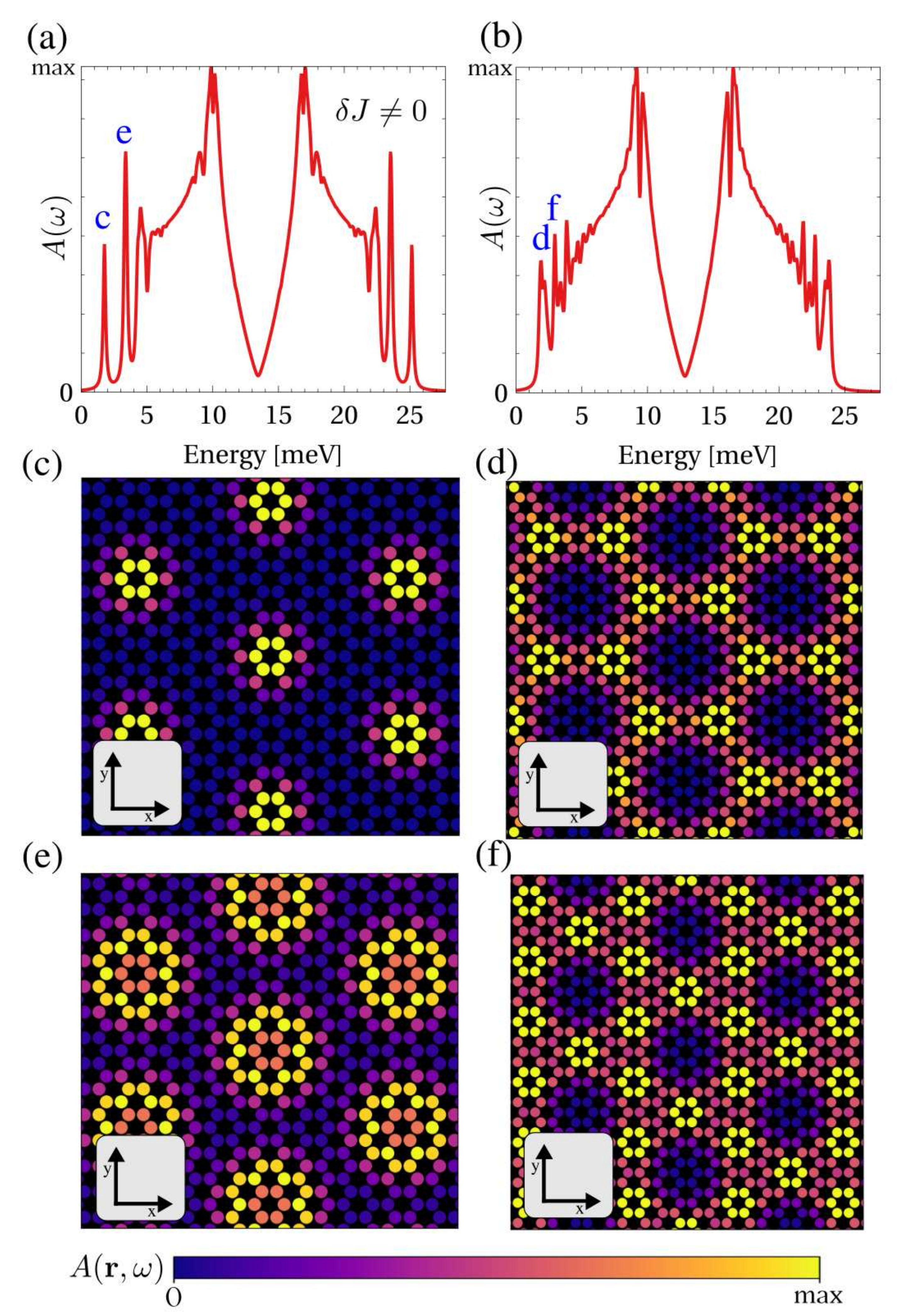}
	\caption{(a,b) Magnon spectral function and (c-f) real-space
	distribution of the modes marked in panels (a,b). It is observed that
	the moir\'e singularities lead to moire modes showing
	a non-uniform distribution in the moir\'e unit cell.}
    \label{figSM:moire_modes}
    %\end{minipage}
    
\end{figure}

\section{Theoretical model}

\subsection{Magnon Hamiltonian}

Here we elaborate on the theoretical model accounting for magnetic excitations
in CrBr$_3$. The magnetic Cr sites realize a $S=3/2$ honeycomb lattice, with
a Heisenberg Hamiltonian of the form

\begin{equation}
    \H = -\sum_{\langle ij \rangle } J_{ij} \vec S_i \cdot \vec S_j 
    -K \sum_{\langle ij \rangle } S^z_i S^z_j 
    -D\sum_{i} \left ( S^z_i \right )^2
    +
    \H_{HO}
    \label{eq:heisenberg}
\end{equation}
where $J_{ij}$ are the isotropic exchange coupling, $K$ the anisotropic exchange, $D$ the single ion anisotropic.
From the fundamental point of view, the terms $K$ and $D$ give rise to a small gap in the magnon gap,
whereas $J_{ij}$ gives rise to the magnon moir\'e bands.
The term $\H_{HO}$ includes additional terms to the Hamiltonian
including
Kitaev exchange\cite{Xu2018}, biquadratic exchange\cite{Kartsev2020}, 
Dzyaloshinskii-Moriya interaction\cite{chen2018topological}, 
and dipolar coupling\cite{Lu2020}. For the sake of concreteness, we will
focus our discussion on the modulation of the exchange coupling
$J_{ij}=J ((\mathbf r_i + \mathbf r_j)/2)$, which is expected
to be the strongest modulation\cite{Soriano2019,Sivadas2018,Chen2019,Song2021}.

We take as starting point the ferromagnetic state of the Hamiltonian Eq.~\ref{eq:heisenberg}.
We derive the magnonic Hamiltonian
using a Holstein-Primakoff transformation\cite{PhysRev.58.1098}
in the low temperature regime as $S^z_n = S -a^\dagger_n a_n$, $S^-_n \approx \sqrt{2S}a^\dagger_n$
and  $S^{+}_n \approx \sqrt{2S}a_n$, with $a^\dagger_n$ $a_n$ creation and annihilation magnon operators
in site $n$. The bosonic magnonic Hamiltonian takes the form

\begin{equation}
    \H = 
    -\sum_{ij}\gamma_{ij} a^\dagger_i a_j 
    + \Delta \sum_i a^\dagger_i a_i+ \text{h.c.}
\end{equation}
where $\gamma_{ij} \sim J_{ij}$ and $\Delta$ controls the magnon gap,
leading to a gapless spectra for $D=K=0$. As a reference, the full magnon
spectra of CrBr$_3$ is approximately $25$ meV\cite{ghazaryan2018magnon},
consistent with first principles
calculations results\cite{Zhang2015}.

\subsection{Moir\'e magnons}

The existence of a moir\'e pattern leads to a modulation of the exchange constants $J_{ij}$, in turn leading to
a modulation of the magnon hoppings $\gamma_{ij}$. From the structural point of view, the spatially dependent
stacking leads to a modulation of the super-exchange
interaction mediated by the substrate. We show
in Fig.~\ref{figSM:moire} the evolution of the magnon spectral function as a function
of the moire length $L_M$ of the modulation (Fig.~\ref{figSM:moire}a,b) and the strength
of the modulation $\delta J$ (Fig.~\ref{figSM:moire}c,d). 
Given the underlying structure of the materials,
we consider a harmonic profile with $C_6$ symmetry taking the form

\begin{equation}
J(\mathbf r) = c_0 + c_1 \sum_{\mathbf G} \cos{ \mathbf G \cdot  \mathbf r}    
\end{equation}
where $\mathbf G$ are the reciprocal lattice vectors of the triangular lattice moire unit cell.
We take by definition 
$\langle J \rangle  = \langle J (\mathbf r) \rangle$, and
$\delta J  = \text{max} [J (\mathbf r)] - \text{min} [J (\mathbf r)]$, and we consider
the two distributions shown in Fig.~\ref{figSM:moire}e,f.
Panels Fig.~\ref{figSM:moire}a,c correspond to the profile Fig.~\ref{figSM:moire}e
(minimum at $\mathbf r = 0$),
and panels Fig.~\ref{figSM:moire}b,d correspond to the profile Fig.~\ref{figSM:moire}f
(maximum at $\mathbf r = 0$). 
It is observed that in both cases moir\'e magnon bands appear at low energies, and with
an energy splitting decreasing as the moir\'e length gets bigger (Fig. \ref{figSM:moire}a,b). 
The splitting between
low energy modes increases as the $\delta J$ modulation becomes stronger, as shown in (Fig.~\ref{figSM:moire}c,d).

The appearance of moir\'e magnons can also be observed in the unfolded magnon structure, as shown in
Fig.~\ref{figSM:unfolded}. In particular, the evolution of the magnon spectral
function with increasing modulation strength shows the appearance of moir\'e mini-bands, which
are ultimately are responsible of the inelastic quasiparticle interference observed experimentally.
This can be observed by computing the isosurfaces of the magnon moir\'e modes, unfolded to the
original moir\'e unit cell as shown in Fig.~\ref{figSM:FS}a,c,e. In particular, it is clearly observed the emergence of
short wavelength features in the whole energy range when moir\'e is switched on. In start
contrast, in the absence of a moire pattern (Fig.~\ref{figSM:FS}b,d,f) the isosurfaces are featureless.
Finally, associated with the magnon reconstruction, the spatial distribution is expected to be affected.
In Fig.~\ref{figSM:moire_modes}, we show the moir\'e magnon spectra for a specific moire strength, for the two profiles
shown in Fig.~\ref{figSM:moire}e,f. It is observed that the low energy moir\'e magnons show a non-uniform distribution in the moir\'e
unit cell (Fig.~\ref{figSM:moire_modes}c-f), featuring emergent triangular, honeycomb and Kagome magnon lattices.

\begin{figure*}[t!]
    \centering
    \includegraphics[width=.95\textwidth]{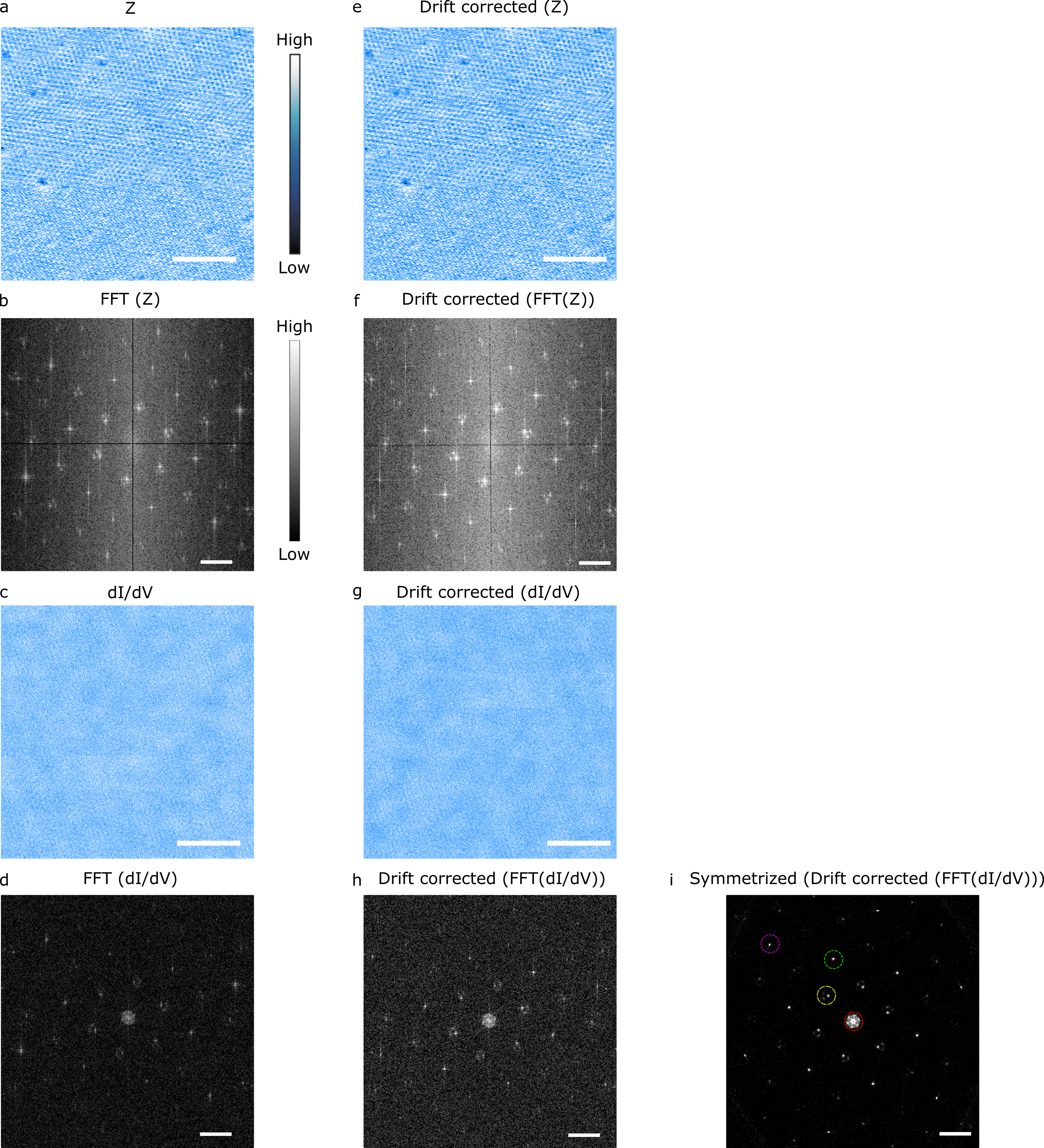}
	\caption{Real space images and fourier transforms after each step: (a),(b) Raw topographic image and its fast fourier transform (FFT), (c), (d) raw tunneling conductance (d$I/$d$V$) and its FFT. Drift corrected (Lawler-Fujita algorithm) (e), (f) topographic image and its FFT, (g), (h) (d$I/$d$V$) and its FFT. (i) 3-fold symmetrised FFT. Constant current map taken at bias V= 2.3 mV, r.m.s modulation=120 $\mu$V  setpoint= 200 pA. The scale bars for real space images (a,c,e,g) 15 nm. The scale bars for reciprocal space images (b, d, f, h, i) are 5 nm$^{-1}$. In (i) red, yellow, green and magenta dotted circles indicate real space length scales of 7 nm, 1.25 nm, 6 {\AA}, 4 {\AA} respectively.}
    \label{figSM:fft}
\end{figure*}

\section{Drift correction and symmetrization of raw images}

The signal to noise ratio of the raw QPI images are enhanced in two steps. At first, the raw constant current topography and d$I/$d$V$ maps are corrected for both piezoelectric mechanical creep and thermal drift by alogorithm developed by Lawler-Fujita  \cite{lawler2010intra}. Finally the drift corrected fourier transform is symmetrized due to the present C$_6$ symmetry of the crystal lattice. The raw, drift corrected and symmetrized images are shown in (Fig.~\ref{figSM:fft}a-i).  

%\clearpage

\bibliography{biblio}